\documentclass[showpacs,preprintnumbers,prd,nofootinbib,floats,amssymb,floatfix]{revtex4}
\usepackage{amsmath}
\usepackage{graphicx}
\usepackage{amsxtra}
\usepackage{hyperref}
\usepackage{amsmath}
\usepackage{amstext}
\begin{document}
\title{The Hamilton-Jacobi characteristic equations  for three dimensional Ashtekar gravity}
\author{Alberto Escalante}  \email{aescalan@ifuap.buap.mx}
 \affiliation{  Instituto de F{\'i}sica, Benem\'erita Universidad Aut\'onoma de Puebla. \\
 Apartado Postal J-48 72570, Puebla Pue., M\'exico, }
 \author{ M. Eduardo Hern\'andez-Garc{\'i}a}  \email{}
 \affiliation{ Facultad de Ciencias F\'{\i}sico Matem\'{a}ticas, Benem\'erita Universidad Au\-t\'o\-no\-ma de Puebla,
 Apartado postal 1152, 72001 Puebla, Pue., M\'exico.}
\begin{abstract}
The Hamilton-Jacobi    analysis of three dimensional  gravity  defined in terms  of Ashtekar-like variables  is performed. We report a detailed analysis where the complete set of Hamilton-Jacobi  constraints,  the characteristic equations and the gauge transformations of the theory are found. We find from  integrability  conditions on the Hamilton-Jacobi   Hamiltonians  that the theory is reduced to a $BF$ field  theory defined  only in terms of  self-dual (or anti-self-dual) variables;  we identify the dynamical variables and the counting of physical degrees of freedom is performed. In addition, we compare our results with those reported by using the canonical formalism. 
\end{abstract}
 \date{\today}
\preprint{}
\maketitle
\section{Introduction}
The study of gauge systems  is the  cornerstone for understand  the fundamental interactions of  nature. Gauge systems are characterized since  there exist   equivalent clases among physical states and they  are connected via  gauge transformations. In fact, the  gauge transformations  are an important part  of the symmetries of the system  because they characterize the core of a gauge theory. The identification of the gauge symmetries  can be carry out by means different and powerful  approaches  such as  the canonical framework developed by Dirac and Bergmann \cite{1, 2}, the symplectic method of Faddeev-Jackiw \cite{3, 4, 5, 6, 7, 8, 9, 10, 11, 12, 13, 14, 15, 16, 17, 18}  and the Hamilton-Jacobi  [HJ] procedure \cite{19, 20, 21, 22}. The $HJ$ approach is an economical and elegant scheme for study   gauge systems;  it is based on the construction of a fundamental differential which has as principal components the $HJ$ constraints called Hamiltonians, which can be   involutives and noninvolutives. The former are  characterized by considering that their   Poisson brackets with all Hamiltonians,  including themselves vanish, in otherwise  they are noninvolutives.  The identification of noninvolutive Hamiltonians allows  us to construct  the so-called generalized brackets which are a generalization of the Poisson brackets,  at the end of the procedure the fundamental differential will be expressed   in terms of involutive Hamiltonians and  the generalized  brackets.  The $HJ$ method has been applied   for studying    several gauge systems,  in particular  systems with general covariance just like $BF$ theories \cite{23, 24}, topological invariants \cite{25} and  field theories  \cite{26, 27}. In this respect, it  has been showed that the development of the $HJ$ scheme is more economical with respect either  Dirac or the FJ approaches.  Furthermore, in the $HJ$ scheme the identification of the symmetries is made  in direct form  and we avoid  the large  procedure  of classification of  constraints just like in Dirac's method is done. In this manner, the $HJ$ framework is an interesting alternative for analyzing gauge systems and their  symmetries. \\
With  the ideas exposed above, in this paper we will apply the $HJ$ approach to 3D gravity. The model under  study is reported  in \cite{28}   and represents   the  3D equivalent version  of  real gravity  theory  reported by  Holst \cite{29}. In fact, the 3D gravity action is expressed in terms of  triads, connexions,  a vector field defined on $R^4$  and a Barbero-Immirzi-like  parameter ($\gamma$).  The analysis reported in \cite{28} was developed by using the canonical scheme and it was showed that the action is reduced to a $BF$ field theory without the presence of the Barbero parameter. In this paper, we will   perform a different analysis by introducing a set of Ashtekar-like variables,  then the $HJ$ formalism will  be developed. We show that the action  is  reduced to a $BF$ field  theory, however the dynamical variables will be identified with the anti-self-dual Ashtekar-like variables such as  it is presented in the Holst paper. On the other hand, the $\gamma$ parameter will be present  through  the  Lagrange multipliers; the  parameter will not contribute to the constraints  because the Lagrange  multipliers are not dynamical, this result represent a difference  with respect to  the Holst action. \\
The paper is organized as follows. In the Sect. II we will introduce the Ashtekar-like variables, then the $HJ$ is performed. The characteristic equations and the symmetries of the theory will be found. In addition, we will observe that the system  is reduced to a $BF$  theory and the Barbero-like parameter will be present  at Lagrange multipliers level. Finally in Sect. III some remarks and conclusions  are exposed.      
\section{Hamilton-Jacobi analysis}
We shall  analyze the following action  \cite{28}
\begin{equation}
\displaystyle{S= \int d^{3}x \epsilon ^{\mu \nu \rho} \left( \frac{1}{2}\epsilon _{IJKL}x^{I}e^{\,\,\, J}_{\mu}F^{\,\,\,\,\,\, KL}_{\nu \rho} + \gamma ^{-1}x_{I}e_{\mu J}F^{\,\,\,\,\,\,IJ}_{\nu \rho} \right)},
\label{eqn:accion-propuesta}
\end{equation}
where  $\epsilon _{IJKL}$ is the volume element of  $SO(4)$, $x^{I}$  is a vector of $R^{4}$, $F^{\,\,\,\,\,\, IJ}_{\mu \nu} = \partial _{\mu} A _{\nu}^{\,\,\, IJ} - \partial _{\nu} A _{\mu}^{\,\,\, IJ} + A_{\mu \,\,\, L}^{\,\,\, I}A_{\nu}^{\,\,\, LJ} - A_{\nu \,\,\, L}^{\,\,\, I}A_{\mu}^{\,\,\, LJ}$ is the strength  curvature of  $SO(4)$, $e^{\,\,\, J}_{\mu}$  is the triad and $ \gamma $   the  Barbero-Immirzi-like  parameter. It is straightforward to prove that  the action (\ref{eqn:accion-propuesta}) describes 3D Euclidean gravity and we can observe that there is a closed relation with the Holst action in the sense that there is a coupling  $\gamma$ parameter. \\
Furthermore, in this paper, we will work with the temporal gauge,  by fixing    $x^{I}=(1,0,0,0)$. Although  there are other gauge fixing options, we are interested  to report the closer relation with the Holst action, and the temporal gauge provides us that aim.  \\ 
In this manner, by performing the 2+1 decomposition the action takes the following form 
\begin{equation}
S = \int d^{3}x \frac{1}{2}\epsilon ^{ab}e_{0i} \left[ \epsilon ^{i}_{\,\,\, jk}F_{ab}^{\,\,\,\,\,\, jk} + \frac{2}{\gamma} F_{ab}^{\,\,\,\,\,\, 0i}  \right] -  \int d^{3}x \epsilon ^{ab}e_{ai} \left[ \epsilon ^{i}_{\,\,\, jk}F_{0b}^{\,\,\,\,\,\, jk} + \frac{2}{\gamma} F_{0b}^{\,\,\,\,\,\, 0i}  \right],
\label{eqn:accion-propuesta-en-la-norma-temporal} 
\end{equation}
where the strength curvature components are given by  
\begin{align}
\nonumber F^{\,\,\,\,\,\, jk}_{a b} &= \partial _{a} A_{b}^{\,\,\, jk} - \partial _{b} A_{a}^{\,\,\, jk} + A_{a \,\,\, 0}^{\,\,\, j}A_{b}^{\,\,\, 0k} - A_{b \,\,\, 0}^{\,\,\, j}A_{a}^{\,\,\, 0k}
+ A_{a \,\,\, l}^{\,\,\, j}A_{b}^{\,\,\, lk} - A_{b \,\,\, l}^{\,\,\, j}A_{a}^{\,\,\, lk}, \\ 
\nonumber F^{\,\,\,\,\,\, 0i}_{a b} &= \partial _{a} A_{b}^{\,\,\, 0i} - \partial _{b} A_{a}^{\,\,\, 0i} + A_{a \,\,\, 0}^{\,\,\, 0}A_{b}^{\,\,\, 0i} - A_{b \,\,\, 0}^{\,\,\, 0}A_{a}^{\,\,\, 0i}
+ A_{a \,\,\, l}^{\,\,\, 0}A_{b}^{\,\,\, li} - A_{b \,\,\, l}^{\,\,\, 0}A_{a}^{\,\,\, li}, \\
 F^{\,\,\,\,\,\, jk}_{0 b} &= \partial _{0} A_{b}^{\,\,\, jk} - \partial _{b} A_{0}^{\,\,\, jk} + A_{0 \,\,\, 0}^{\,\,\, j}A_{b}^{\,\,\, 0k} - A_{b \,\,\, 0}^{\,\,\, j}A_{0}^{\,\,\, 0k} 
 + A_{0 \,\,\, l}^{\,\,\, j}A_{b}^{\,\,\, lk} - A_{b \,\,\, l}^{\,\,\, j}A_{0}^{\,\,\, lk} , \label{eqn:componentes-de-la-curvatura} \\
 \nonumber  \\ F^{\,\,\,\,\,\, 0i}_{0 b} &= \partial _{0} A_{b}^{\,\,\, 0i} - \partial _{b} A_{0}^{\,\,\, 0i} + A_{0 \,\,\, 0}^{\,\,\, 0}A_{b}^{\,\,\, 0i} - A_{b \,\,\, 0}^{\,\,\, 0}A_{0}^{\,\,\, 0i} 
 + A_{0 \,\,\, l}^{\,\,\, 0}A_{b}^{\,\,\, li} - A_{b \,\,\, l}^{\,\,\, 0}A_{0}^{\,\,\, li}. 
\end{align}
Now we introduce the following Ashtekar-like  variables 
\begin{equation}
{}^{{}^{\pm}}\mathcal{A}_{b}^{\,\,\, i} =  \epsilon ^{i}_{\,\,\, jk}A_{b}^{\,\,\, jk} \pm \frac{2}{\gamma}A_{b}^{\,\,\, i0}.
\label{eqn:variables-tipo-ashtekar} 
\end{equation}
and thus we  obtain  
\begin{align}
A_{b}^{\,\,\, i0} &= \frac{\gamma}{4} \left[ {}^{{}^{+}}\mathcal{A}_{b}^{\,\,\, i} - {}^{{}^{-}}\mathcal{A}_{b}^{\,\,\, i} \right], \\
A_{b}^{\,\,\, jk} &= \frac{1 }{4} \epsilon ^{jk}_{\,\,\,\,\,\, i} \left[ {}^{{}^{+}}\mathcal{A}_{b}^{\,\,\, i} + {}^{{}^{-}}\mathcal{A}_{b}^{\,\,\, i} \right].
\label{eqn:inversa-de-las-variables-tipo-ashtekar} 
\end{align}
Then, by using these variables,  the action reads 
{\scriptsize { 
\begin{align}
S = \int d^{3}x \epsilon ^{ab}e_{0i} \left[ \partial _{a}{}^{{}^{-}}\mathcal{A}_{b}^{\,\,\, i} - \epsilon ^{i}_{\,\,\, jk}
\nonumber \left[ \left(\frac{\gamma ^{2}-1}{16} \right){}^{{}^{+}}\mathcal{A}_{a}^{\,\,\, j}{}^{{}^{+}}\mathcal{A}_{b}^{\,\,\, k} + \left(\frac{\gamma ^{2}+3}{16} \right){}^{{}^{-}}\mathcal{A}_{a}^{\,\,\, j}{}^{{}^{-}}\mathcal{A}_{b}^{\,\,\, k} - \left(\frac{\gamma ^{2}-1}{8}\right){}^{{}^{+}}\mathcal{A}_{a}^{\,\,\, j}{}^{{}^{-}}\mathcal{A}_{b}^{\,\,\, k} \right] \right] \\
+ \int d^{3}x \epsilon ^{ab}e_{ai} \left[ \partial _{b}\left( \epsilon ^{i}_{\,\,\, jk}A_{0}^{\,\,\, jk} + \frac{2}{\gamma}A_{0}^{\,\,\, 0i}  \right) - \partial _{0} {}^{{}^{-}}\mathcal{A}_{b}^{\,\,\,i} + \left(\frac{\gamma ^{2}-1}{2 \gamma} \right)\epsilon ^{i}_{\,\,\, jk} A_{0}^{\,\,\, j0}{}^{{}^{+}}\mathcal{A}_{b}^{\,\,\, k} - \left[ \left(\frac{\gamma ^{2}+1}{2 \gamma} \right)\epsilon ^{i}_{\,\,\, jk}A_{0}^{\,\,\, j0} + A_{0 \,\,\, k}^{\,\,\, i} \right] {}^{{}^{-}}\mathcal{A}_{b}^{\,\,\, k} \right].
\label{eqn:accion-con-variables-tipo-ashtekar} 
\end{align}}}
we can observe that if $\gamma=1$, then the self-dual-conexion disapears. Because of  the action is under a variational principle, we will not  fix the value of $\gamma$ until the end of the calculations. In this manner, according the $HJ$ method,  from the definition of the momenta  $(p^0_i, p^a_i, \pi_i,  \pi_{ij}, {{}^{+}}\pi^a_i, {{}^{-}}\pi^a_i)$ canonically conjugated  to $(e_0 ^i, e_a^i, A_{0}^{\,\,\, i0}, A_{0}^{\,\,\, ij},  {^{+}}\mathcal{A}_{a}^{\,\,\, i},  {^{-}}\mathcal{A}_{a}^{\,\,\, i})$,   and from the  action (\ref{eqn:accion-con-variables-tipo-ashtekar}) we identify the following Hamiltonians 
\begin{eqnarray}
 H'&\equiv&\pi + H_0=0, \nonumber \\
 \phi_i&\equiv& p^0_i=0, \nonumber \\
  \phi^a_i&\equiv&p^a_i =0, \nonumber \\
  \tilde{\phi}_i &\equiv&  \pi_i=0, \nonumber \\
  \phi_{ij} &\equiv&  \pi_{ij}=0, \nonumber \\
   {{}^{+}}\phi^a_i &\equiv&  {{}^{+}}\pi^a_i =0, \nonumber \\ 
  {{}^{-}} \phi^a_i&\equiv& {{}^{-}}\pi^a_i - \epsilon^{ab} e_{bi}=0, 
\end{eqnarray}
with $\pi= \partial_0S$ where $S$ is the action and the canonical Hamiltonian $H_0$ reads
{\scriptsize { 
\begin{align}
H_0=-\epsilon ^{ab}e_{0i} \left[ \partial _{a}{}^{{}^{-}}\mathcal{A}_{b}^{\,\,\, i} - \epsilon ^{i}_{\,\,\, jk}
\nonumber \left[ \left(\frac{\gamma ^{2}-1}{16} \right){}^{{}^{+}}\mathcal{A}_{a}^{\,\,\, j}{}^{{}^{+}}\mathcal{A}_{b}^{\,\,\, k} + \left(\frac{\gamma ^{2}+3}{16} \right){}^{{}^{-}}\mathcal{A}_{a}^{\,\,\, j}{}^{{}^{-}}\mathcal{A}_{b}^{\,\,\, k} - \left(\frac{\gamma ^{2}-1}{8}\right){}^{{}^{+}}\mathcal{A}_{a}^{\,\,\, j}{}^{{}^{-}}\mathcal{A}_{b}^{\,\,\, k} \right] \right] \\
+ {{}^{-}} \pi^b_i\left[ \partial _{b}\left( \epsilon ^{i}_{\,\,\, jk}A_{0}^{\,\,\, jk} + \frac{2}{\gamma}A_{0}^{\,\,\, 0i}  \right) + \left(\frac{\gamma ^{2}-1}{2 \gamma} \right)\epsilon ^{i}_{\,\,\, jk} A_{0}^{\,\,\, j0}{}^{{}^{+}}\mathcal{A}_{b}^{\,\,\, k} - \left[ \left(\frac{\gamma ^{2}+1}{2 \gamma} \right)\epsilon ^{i}_{\,\,\, jk}A_{0}^{\,\,\, j0} + A_{0 \,\,\, k}^{\,\,\, i} \right] {}^{{}^{-}}\mathcal{A}_{b}^{\,\,\, k} \right].
\end{align}}}
Now with the Hamiltonians we construct the following fundamental $HJ$ differential \cite{19, 20, 21, 22, 23, 24, 25}
\begin{eqnarray}
df(x) &=&\int d^{3}y\Big(\{f(x), H'\}dt + \{f(x), \phi_i \}d\xi{^i} + \{f(x), \phi^a_i \}d\xi_a{^i} + \{f(x), \tilde{\phi}_i \}d\tilde{\xi}^{i} + \{f(x), \phi_{ij} \}d\xi{^{ij}}  \nonumber \\
&+& \{f(x), {{}^{+}}\phi^a_i \}d{^{+}}\xi_a{^i}  +\{f(x), {{}^{-}} \phi^a_i  \}d{^{-}}\xi_a{^i}\Big),
\end{eqnarray}
where $(\xi{^i}, \xi_a{^i}, \tilde{\xi}^{i}, \xi{^{ij}}, {^{+}}\xi_a{^i}, {^{-}}\xi_a{^i} )$ are parameters related to the Hamiltonians. It is worth to mention that these parameters play a fundamental roll;  for  involutives Hamiltonians they  correspond to  parameters  related with  the gauge transformations, this fact will be discussed bellow.  \\
On the other hand, the fundamental Poisson brackets between the canonical variables are given by 
\begin{eqnarray}
\{ e_\mu ^i(x), p^\alpha_j(y)  \}&=&\delta^\alpha_\mu \delta^i_ j\delta^2(x-y), \nonumber \\
\{ A_{0}^{\,\,\, i0}(x) ,  \pi_j(y)\}&=& \delta{^i}_j \delta^2(x-y), \nonumber \\
\{ A_{0}^{\,\,\, ij} (x), \pi_{kl}(y) \}&=& \frac{1}{2}\delta^{ij}_{kj} \delta^2(x-y), \nonumber \\
\{ {^{+}}\mathcal{A}_{a}^{\,\,\, i} (x), {{}^{+}}\pi^b_j (y) \}&=& \delta{^b}_a \delta{^i}_j \delta^2(x-y), \nonumber \\
\{ {^{-}}\mathcal{A}_{a}^{\,\,\, i} (x), {{}^{-}}\pi^b_i  (y)\}&=& \delta{^b}_a \delta{^i}_j \delta^2(x-y). 
\label{bra}
\end{eqnarray}
Once defined the Poisson brackets,  all Hamiltonians having vanishing Poisson brackets to each other  are called   involutives, otherwise, they are non-involutive Hamiltonians. Thus, by using the fundamental brackets we observe that the Hamiltonians $(\phi_i, \tilde{\phi}_i , \phi_{ij},  {{}^{+}}\phi^a_i )$ are  involutives and 
$ (\phi^a_i, {{}^{-}} \phi^a_i )$ are noninvolutives. Furthermore, due to  there are noninvolutives Hamiltonians, we introduce the generalized brackets  by constructing the matrix whose entries are the Poisson brackets between  all noninvolutives  Hamiltonians, this is
\begin{eqnarray*}
\label{eq}
C_{\alpha\beta}=
\left(
  \begin{array}{cccc}
  0	&\quad		-\epsilon^{ab}\eta_{ij}\\
\epsilon^{ba}\eta_{ij}	&\quad		0\\
 \end{array}
\right) \delta^{2}(x-y),
\end{eqnarray*}

and its inverse reads 

\begin{eqnarray*}
\label{eq}
(C{_{\alpha\beta}})^{-1}=
\left(
  \begin{array}{cccc}
 0	&\quad		\epsilon_{ad}\eta^{ij}\\
-\epsilon_{bd}\eta^{ij}	&\quad		0	\\
 \end{array}
\right) \delta^{2}(x-y),
\end{eqnarray*}
thus, by using $(C{_{\alpha\beta}})^{-1}$  we can introduce   the generalized brackets  given by \cite{21, 22, 23, 24}
\begin{align}
\{A, B\}^{*} &=\{A, B\} - \{A, H'_{\bar{a}}\}(C{_{\bar{a}\bar{b}}})^{-1}\{H'_{\bar{b}}, B\},
\label{10}
\end{align}
where $H'_{\bar{a}}$ are the non-involutive Hamiltonians. In this manner, by using (\ref{10}) the  generalized brackets are given by
\begin{eqnarray}
\{ e_0 ^i(x), p^0_j(y)  \}^{*}&=& \delta^i_ j\delta^2(x-y), \nonumber \\
\{ e_a ^i(x), p^b_j(y)  \}^{*}&=&0, \nonumber \\
\{ A_{0}^{\,\,\, i0}(x) ,  \pi_j(y)\}^{*}&=& \delta{^i}_j \delta^2(x-y), \nonumber \\
\{ A_{0}^{\,\,\, ij} (x), \pi_{kl}(y) \}^{*}&=& \frac{1}{2}\delta^{ij}_{kj} \delta^2(x-y), \nonumber \\
\{ {^{+}}\mathcal{A}_{a}^{\,\,\, i} (x), {{}^{+}}\pi^b_j (y) \}^{*}&=& \delta{^b}_a \delta{^i}_j \delta^2(x-y), \nonumber \\
\{ {^{-}}\mathcal{A}_{a}^{\,\,\, i} (x), {{}^{-}}\pi^a_i  (y)\}^{*}&=& \delta{^b}_a \delta{^i}_j \delta^2(x-y). 
\label{bra}
\end{eqnarray}
The introduction of the generalized brackets redefine the dynamics. In fact,  the non-involutive constraints are removed  from the fundamental differential and  it can be   expressed  in terms of the generalized brackets and involutive Hamiltonians. \\
In this manner, the fundamental differential written  in terms of the generalized brackets and involutive  Hamiltonians takes  the form
%%%%%%%%%%%
\begin{eqnarray}
df(x) &=&\int d^{3}y\Big(\{f(x), H'\}^{*}dt + \{f(x), \phi_i \}^{*}d\xi{^i}  + \{f(x), \tilde{\phi}_i \}^{*}d\tilde{\xi}^{i} + \{f(x), \phi_{ij} \}^{*}d\xi{^{ij}}  \nonumber \\
&+& \{f(x), {{}^{+}}\phi^a_i \}^{*}d{^{+}}\xi_a{^i} \Big),
\end{eqnarray}
thus,  the \textit{Frobenius integrability} conditions for the Hamiltonians \cite{19, 20}, say  $(\phi_i, \tilde{\phi}_i , \phi_{ij},  {{}^{+}}\phi^a_i )$, introduce new Hamiltonians 
\begin{eqnarray}
d\phi_i, &=&\int d^{3}y\Big(\{\phi_i, H'\}^{}dt + \{\phi_i, \phi_j \}^{}d\xi{^j}  + \{\phi_i, \tilde{\phi}_j \}^{}d\tilde{\xi}^{j} + \{\phi_i, \phi_{kl} \}^{}d\xi{^{kl}} +  \{\phi_i, {{}^{+}}\phi^a_j \}^{}d{^{+}}\xi_a{^j} \Big) \nonumber  \\
&=& \epsilon^{ab}\left[ \partial _{a}{}^{{}^{-}}\mathcal{A}_{b}^{\,\,\, i} - \epsilon ^{i}_{\,\,\, jk}
\nonumber \left[ \left(\frac{\gamma ^{2}-1}{16} \right){}^{{}^{+}}\mathcal{A}_{a}^{\,\,\, j}{}^{{}^{+}}\mathcal{A}_{b}^{\,\,\, k} + \left(\frac{\gamma ^{2}+3}{16} \right){}^{{}^{-}}\mathcal{A}_{a}^{\,\,\, j}{}^{{}^{-}}\mathcal{A}_{b}^{\,\,\, k} - \left(\frac{\gamma ^{2}-1}{8}\right){}^{{}^{+}}\mathcal{A}_{a}^{\,\,\, j}{}^{{}^{-}}\mathcal{A}_{b}^{\,\,\, k} \right] \right] =0, \nonumber  \\
d\tilde{\phi}_i , &=&\int d^{3}y\Big(\{\tilde{\phi}_i , H'\}^{}dt + \{\tilde{\phi}_i , \phi_j \}^{}d\xi{^j}  + \{\tilde{\phi}_i  \tilde{\phi}_j \}^{}d\tilde{\xi}^{j} + \{\tilde{\phi}_i , \phi_{kl} \}^{}d\xi{^{kl}} +  \{\tilde{\phi}_i , {{}^{+}}\phi^a_j \}^{}d{^{+}}\xi_a{^j} \Big)\nonumber  \\
&=&  \frac{2}{\gamma}\partial _{a}{{}^{-}} \pi^a_i+ \epsilon ^{j}_{\,\,\, ik} {{}^{-}} \pi^a_j\left[ \left(\frac{\gamma ^{2}-1}{2 \gamma} \right) {}^{{}^{+}}\mathcal{A}_{a}^{\,\,\, k} - \left(\frac{\gamma ^{2}+1}{2 \gamma} \right) {}^{{}^{-}}\mathcal{A}_{a}^{\,\,\, k} \right] = 0,   \nonumber  \\
d\phi_{ij}, &=&\int d^{3}y\Big(\{\phi_{ij}, H'\}^{}dt + \{\phi_{ij}, \phi_k \}^{}d\xi{^k}  + \{\phi_{ij}, \tilde{\phi}_k \}^{}d\tilde{\xi}^{k} + \{\phi_{ij}, \phi_{kl} \}^{*}d\xi{^{kl}} +  \{\phi_{ij}, {{}^{+}}\phi^a_k\}^{}d{^{+}}\xi_a{^k} \Big) \nonumber  \\
&=&  \epsilon ^{k \,\,\,\,\, }_{\,\,\, ij}\partial _{a}{{}^{-}} \pi^a_k + \frac{1}{2}({{}^{-}} \pi^a_i{}^{{}^{-}}\mathcal{A}_{aj}- {{}^{-}} \pi^a_j{{}^{-}}\mathcal{A}_{ai})=0 \rightarrow  \epsilon{_{ij}}^lD_a{{}^{-}}\pi^a_l=0,  \nonumber  \\
d{{}^{+}}\phi^a_i , &=&\int d^{3}y\Big(\{{{}^{+}}\phi^a_i , H'\}^{}dt + \{{{}^{+}}\phi^a_i , \phi_j \}^{}d\xi{^j}  + \{{{}^{+}}\phi^a_i , \tilde{\phi}_j \}^{}d\tilde{\xi}^{j} + \{{{}^{+}}\phi^a_i , \phi_{kl} \}^{}d\xi{^{kl}} +  \{{{}^{+}}\phi^a_i , {{}^{+}}\phi^a_j \}^{}d{^{+}}\xi_a{^j} \Big) \nonumber  \\
&=& \frac{1}{4} \epsilon ^{ab}\epsilon ^{j}_{\,\,\, ik}e_{0j} \left( {}^{{}^{+}}\mathcal{A}_{b}^{\,\,\, k} - {}^{{}^{-}}\mathcal{A}_{b}^{\,\,\, k} \right) - \frac{1}{\gamma} \epsilon ^{j}_{\,\,\, ik} {{}^{-}} \pi^a_jA_{0}^{\,\,\, k0} = 0,  
\end{eqnarray}
%%%%%%%%%%%%%%%%%%%%%%%%%
where  we identify  the following   Hamiltonians, 
\begin{eqnarray}
\chi^{i} &\equiv& \epsilon^{ab}\left[ \partial _{a}{}^{{}^{-}}\mathcal{A}_{b}^{\,\,\, i} - \epsilon ^{i}_{\,\,\, jk}
\nonumber \left[ \left(\frac{\gamma ^{2}-1}{16} \right){}^{{}^{+}}\mathcal{A}_{a}^{\,\,\, j}{}^{{}^{+}}\mathcal{A}_{b}^{\,\,\, k} + \left(\frac{\gamma ^{2}+3}{16} \right){}^{{}^{-}}\mathcal{A}_{a}^{\,\,\, j}{}^{{}^{-}}\mathcal{A}_{b}^{\,\,\, k} - \left(\frac{\gamma ^{2}-1}{8}\right){}^{{}^{+}}\mathcal{A}_{a}^{\,\,\, j}{}^{{}^{-}}\mathcal{A}_{b}^{\,\,\, k} \right] \right] =0, \nonumber \\
\tilde{\chi}_i   &\equiv&   \frac{2}{\gamma}\partial _{a}{{}^{-}} \pi^a_i+ \epsilon ^{j}_{\,\,\, ik} {{}^{-}} \pi^a_j\left[ \left(\frac{\gamma ^{2}-1}{2 \gamma} \right) {}^{{}^{+}}\mathcal{A}_{a}^{\,\,\, k} - \left(\frac{\gamma ^{2}+1}{2 \gamma} \right) {}^{{}^{-}}\mathcal{A}_{a}^{\,\,\, k} \right] = 0, \nonumber \\
\tilde{\psi}_{ij} &\equiv& \epsilon{_{ij}}^l D_a{{}^{-}}\pi^a_l=0, \nonumber \\
{{}^{+}}\chi^a_i &\equiv&  \frac{1}{4} \epsilon ^{ab}\epsilon ^{j}_{\,\,\, ik}e_{0j} \left( {}^{{}^{+}}\mathcal{A}_{b}^{\,\,\, k} - {}^{{}^{-}}\mathcal{A}_{b}^{\,\,\, k} \right) - \frac{1}{\gamma} \epsilon ^{j}_{\,\,\, ik} {{}^{-}} \pi^a_jA_{0}^{\,\,\, k0} = 0,
\label{16}
\end{eqnarray}
%%%%%%%%%%%%%%%%%%%%%%%%%%%%%%%
here $D_a{{}^{-}}\pi^a_i= \partial_a {{}^{-}}\pi^a_i -\frac{1}{2} \epsilon_{ij}{^k} {}^{{}^{-}}\mathcal{A}_{b}^{\,\,\, j}{{}^{-}}\pi^a_k$. On the other hand, from the  Hamiltonian   ${{}^{+}}\chi^a_i$  we observe that ${{}^{+}}\chi^a_i {{}^{-}} \pi^{ci}+ {{}^{+}}\chi^c_i {{}^{-}} \pi^{ai}=0$, and thus  $ {}^{{}^{+}}\mathcal{A}_{b}^{\,\,\, k} - {}^{{}^{-}}\mathcal{A}_{b}^{\,\,\, k}=0$; this result implies that in three dimensional Ashtekar gravity the dynamical variables are given by the adjoint representation of $SO(3)$. In order to  follow the Holst work \cite{29} we choose  $A_{b}^{\,\,\, jk} = \frac{1 }{4} \epsilon^{jk}_{\,\,\,\,\,\, i}  {}^{{}^{-}}\mathcal{A}_{b}^{\,\,\, i}$. In this manner, with these results at hand the Hamiltonians (\ref{16}) take the form
\begin{align}
\nonumber \mathcal{H}^i:= \frac{1}{2}\epsilon ^{ab} \mathcal{F}^{\,\,\,\,\,\, i} _{ab} = \frac{1}{2}\epsilon ^{ab} \left[ \partial _{a} {}^{{}^{-}}\mathcal{A}_{b}^{\,\,\,i} - \partial _{b} {}^{{}^{-}}\mathcal{A}_{a}^{\,\,\,i}  - \frac{1}{2}\epsilon ^{i}_{\,\,\, jk}{}^{{}^{-}}\mathcal{A}_{a}^{\,\,\, j}{}^{{}^{-}}\mathcal{A}_{b}^{\,\,\, k} \right] = 0, \\
\nonumber \mathcal{G}_{j} := \frac{2}{\gamma}\partial _{b} {}^{{}^{-}}\pi ^{b}_{\,\,\, j} - \frac{1}{\gamma}\epsilon ^{i}_{\,\,\, jk} {}^{{}^{-}}\pi ^{b}_{\,\,\, i}{}^{{}^{-}}\mathcal{A}_{b}^{\,\,\, k} = 0, \\
 \mathcal{\tilde{G}}_{ij} := \epsilon{_{ij}}^lD_a{{}^{-}}\pi^a_l=0,
\label{17}  
\end{align}
and the Hamiltonian $H_0$  is reduced to 
\begin{equation}
H_0= -\frac{\epsilon^{ab}}{2} e_{0i} F^i_{ab}-\Lambda^iD_b {}^{{}^{-}}\pi ^{b}_{\,\,\, i},    
\end{equation}
where $\Lambda^i= \epsilon^i{_{jk}} A_0{^{jk}} - \frac{2}{\gamma} A_0{^{i0}}$. We can observe that the contribution of the  $\gamma$ parameter is only present in $\Lambda^i$ that will be identified as   Lagrange multipliers, this result is a difference respect to   that reported in \cite{28},  where the Barbero-like parameter is eliminated completely. Furthermore,  from Eq. (\ref{17}) we observe that the variables $\pi_{ij}$ and $\pi_i$ generate the same involutive Hamiltonian $\mathcal{G}_{i}$, however we will not remove that Hamiltonian  until the end of the analysis.  The Hamiltonians (\ref{17}) are involutives; their generalized  algebra is closed  
\begin{eqnarray}
\{\mathcal{H}^i (x), \mathcal{H}^j(y)\}&=&0,  \nonumber \\
\{ \mathcal{H}^i (x),  \mathcal{G}_{j} (y) \} &=& - \frac{1}{2} \epsilon^{i}{_{jk}} \mathcal{H}^k, \nonumber \\
\{  \mathcal{G}_{i}(x),  \mathcal{G}_{j}(y) \}&=&  - \frac{1}{2} \epsilon_{ij} {^{k}}  \mathcal{G}_k, 
\end{eqnarray}
because the algebra is closed, then there are not more Hamiltonians. \\ 
Thus, by using all involutive Hamiltonians  we construct a new fundamental  differential   
\begin{eqnarray}
df(x) &=&\int d^{3}y\Big(\{f(x), H_0(y)\}^{*}dt + \{f(x), \phi_i (y)\}^{*}d\xi{^i}  + \{f(x), \tilde{\phi}_i (y)\}^{*}d\tilde{\xi}^{i} + \{f(x), \phi_{ij}(y) \}^{*}d\xi{^{ij}}  \nonumber \\
&+& \{f(x), \mathcal{H}(y) \}^* d\xi + \{f(x), \mathcal{G}_{i} (y) \}^* d\omega^i  + \{f(x), \mathcal{\tilde{G}}_{i}{^k} (y) \}^* d\omega^i{_{k}}  \Big),
\end{eqnarray}
where all noninvolutive Hamiltonians have  been removed.  Now, from the fundamental differential we can identify the characteristics equations,  then the symmetries. The characteristic equations are given by 
\begin{eqnarray}
d {}^{{}^{-}}\mathcal{A}_{a}^{\,\,\,i} &=& D_a \Lambda^i dt +D_a d\tilde{\omega}^i\nonumber\\
d {}^{{}^{-}}\pi ^{a}_{\,\,\, i}&=& \big[ \epsilon^{ab} D_b e_{oi}- \frac{\Lambda^k}{2} \epsilon_{ki}{^{j}}{{}^{-}}\pi ^{a}_{\,\,\, j} \big]dt  + \frac{\epsilon_{ki}{^{j}}}{2} {{}^{-}}\pi ^{a}_{\,\,\, j}d\tilde{\omega}^k ,  \nonumber\\
d p^0_i&=& \frac{\epsilon^{ab}}{2}  F_{iab} dt =0,  \nonumber\\
d\pi_i&=& \mathcal{G}_{i} dt=0, \nonumber\\
d\pi_{ij} &=& \epsilon_{ij}{^k \mathcal{G}_{k} }dt= 0,
\label{21}
\end{eqnarray}
\begin{eqnarray}
d e_0^i&=& d\xi{^i},  \nonumber \\
d A_0{^{ij}}&=& d\xi{^{ij}}, \nonumber \\
d A_0{^{i0}} &=& d\tilde{\xi}^{i},  
\label{22}
\end{eqnarray} 
where we defined  $d\tilde{\omega}^i\equiv d\omega^i + \epsilon^i{_{jk}}d\omega^{jk}  $. From (\ref{21}) we can identify the equations of motion 
\begin{eqnarray}
\partial_0 {{}^{-}}{\mathcal{A}}_{a}^{\,\,\,i} &=& D_a \Lambda^i,  \nonumber \\
\partial_0 {}^{{}^{-}}\pi ^{a}_{\,\,\, i}&=& \big[ \epsilon^{ab} D_b e_{0i}- \frac{\Lambda^k}{2} \epsilon_{ki}{^{j}}{{}^{-}}\pi ^{a}_{\,\,\, j} \big], 
\label{mot}
\end{eqnarray}
and we  observe that the evolution of $p^0_i, \pi_i, \pi_{ij}$ due to the noninvolutive Hamltonians  vanishes. Furthermore, Eq.  (\ref{22}),   implies  that $e_0^i,  A_0{^{ij}}, A_0{^{i0}} $ are identified as Lagrange multipliers, thus,  $\Lambda^i$ is also identified as Lagrange multiplier. Moreover, by taking $dt=0$ in (\ref{21}) we can also identify the gauge transformations of the theory 
\begin{eqnarray}
\delta {}^{{}^{-}}\mathcal{A}_{a}^{\,\,\,i}  &=&D_a \varepsilon^i,  \nonumber \\
\delta {}^{{}^{-}}\pi ^{a}_{\,\,\, i} &=& \frac{1}{2}\epsilon_{ki}{^{j}} {{}^{-}}\pi ^{a}_{\,\,\, j}\varepsilon^k,  
\end{eqnarray}
where $\varepsilon^i \equiv d\tilde{\omega}^i$.  On the other hand, we commented above that the variables   $A_0{^{jk}}$ and $A_0{^{i0}}$ generate the same Hamiltonian $\mathcal{G}_{i}$ and this fact will be taken into account to perform the counting of physical degrees of freedom. In fact, the counting of physical degrees of freedom is performed as follows: there are 12 dynamical variables $ ({{}^{-}}{\mathcal{A}}_{a}^{\,\,\,i}  ,{{}^{-}}\pi ^{a}_{\,\,\, i})$,  and 18 involutive Hamiltonians $(\mathcal{H}^i, \mathcal{G}_{i}, \mathcal{\tilde{G}}_{ij}, p^0_i,  \pi_i, \pi_{ij})$, however, $   \pi_i$ and $ \pi_{ij}$ generate the same hamiltonian $\mathcal{G}_{i}$; therefore there are  18-6=12 independent involutive Hamiltonians, hence the system is laking of  physical  degrees of freedom as expected;  in the counting of degrees of freedom  only independent involutive Hamiltonians must be involved. Moreover, if we fix  $A_0{^{jk}}=0$, we still have the presence of the $\gamma$ parameter in the Lagrange multiplier $\Lambda^i$; the theory will take  a $BF$ form with  $\gamma$  present only at the  level of Lagrange multipliers  and there will be a contribution in the equation of motion (\ref{mot}). On the other hand,  if we remove $\pi_i$ by taking $A_0{^{i0}}=0$, then there is not any contribution from $\gamma$ because $\Lambda^i= \epsilon^i{_{jk}} A_0{^{jk}} $;  the theory will be a $BF$ theory just like that reported in \cite{28},  and hence our results extend the results  reported in the literature. \\
%%%%%%%%%
\section{Conclussions}
In this paper,  the HJ analysis of  3D gravity written in terms of Ashtekar-Like variables was performed. We obtained an action with a close relation  to the Holst Lagrangian where self-dual and anti-self-dual connexions are present. We identified all $HJ$ Hamiltonians of the theory,  then a fundamental differential was constructed. From the fundamental differential the characteristic equations were found; we reported  the gauge transformations and we identified the dynamical variables corresponding to the adjoint representation of the $SO(3)$ conexion. We observed that the theory is reduced to a $BF$ theory where the presence of the Barbero-like parameter is present at level of the Lagrange multipliers, in particular, under a fixing gauge on the multipliers the results reported in the literature were reproduced. It is worth to mention  that the coupling of 3D gravity with degrees of freedom as for instance matter degrees of freedom, will be an interesting scenario  to analize. In fact, the coupling with matter degrees of freedom could provide us  an understanding of the rol of the  $\gamma$  parameter in 3D gravity just like it is present in the 4D case \cite{30}. In this sense, it is well-known  that in 4D gravity with the coupling of fermions, there is a contribution of the $\gamma$ parameter; it does not vanish   and   it determines the coupling constant of a four-fermion interaction.   In this manner,  the $HJ$ framework will be a  good alternative for analyzing these problems   and we expect to find in the future     advantages  in relation with other approaches.
\\
\noindent \textbf{Acknowledgements}\\[1ex]
 We would like to  thank R. Cartas-Fuentevilla for reading  the manuscript.

 \end{document}